\documentclass[prb,twocolumn]{revtex4}

\usepackage[dvips]{graphics}
\usepackage{dcolumn}
\usepackage{bm}

\begin{document}

\title{Photoemission studies of the annealing induced modifications of (GaMn)As}

\author{M. Adell}
\affiliation{Department of Experimental Physics, Chalmers University of Technology, SE-412 96 G\"oteborg, Sweden}

\author{L. Ilver}
\affiliation{Department of Experimental Physics, Chalmers University of Technology, SE-412 96 G\"oteborg, Sweden}

\author{J. Kanski}
\affiliation{Department of Experimental Physics, Chalmers University of Technology, SE-412 96 G\"oteborg, Sweden}

\author{J. Sadowski}
\affiliation{MAX-lab, Lund University, SE-221 00 Lund, Sweden}
\affiliation{ Institute of Physics, Polish Academy of Sciences, al. Lotnikow 32/46, 02-668 Warsaw, Poland}

\author{R. Mathieu}
\affiliation{Department of Materials Science, Uppsala University, SE-751 21 Uppsala Sweden}

\date{\today}

\begin{abstract}
Using angle resolved photoemission we have investigated annealing-induced changes in $Ga_{1-x}Mn_{x}As$ with $x=0.05$. We find that the position of the Fermi energy is a function of annealing time and temperature. It is also established that the Curie temperature is strongly correlated to the separation between the Fermi level and the valence band maximum. Valence band photoemission shows that the Mn3d spectrum is modified by the annealing treatments. 
\end{abstract}

\maketitle

The III-V ferromagnetic semiconductor (GaMn)As has attracted much attention over recent years, both due to possible applications integrating  magnetic and electronic properties, and because of the scientific challenge posed by its unusual properties. One parameter of great interest is the Curie temperature ($T_c$). It is desirable, in particular when considering applications, to be able to control and increase $T_c$  above room temperature. Typical $T_c$ values for as-grown (GaMn)As are in the range 30-80 K. Several factors which influence $T_c$ have been identified, e.g. the Mn concentration \cite{Asklund2}, point defects and the (GaMn)As layer thickness \cite{Sorensen}.  In particular point defects have been a subject of increasing interest, as their limiting influence on ferromagnetism in (GaMn)As has been recognized from post-growth annealing experiments \cite{Hayashi}. Thus, by optimal annealing (with respect to time and temperature), the $T_c$ can be increased from below 100 K to around 150 K \cite{Ku}. Excessive annealing results in deteriorating ferromagnetism. The aim of this study was to investigate how the electronic properties of (GaMn)As are modified by annealing. In particular the Fermi energy ($E_F$) and the valence band (VB) have been studied as functions of annealing and the results are directly related to the modifications of the Curie temperature.
\\
The experiments were performed at the MAX I electron storage ring at the Swedish National Synchrotron Radiation Center MAX-lab, using BL41. The end station at this beamline includes an on-line MBE system, a particularly important asset for the present study since it allows the samples to be transferred from the MBE to the photoemission chamber under UHV conditions. Thus the samples can be investigated without any surface cleaning or other preparation treatments. During the transfer between the growth and analysis chambers  the pressure was in the range of $10^{-9}$ torr, while in the photoemission chamber the pressure was in the low $10^{-10}$ torr range. The MBE system was equipped with five Knudsen cells and a valved $As_{2}$ cracker source. A 10 keV RHEED system was used to monitor the growth. The (GaMn)As layers were grown on 10x10 mm$^2$ n-type GaAs substrates, which were In-glued on Mo-blocks. The sample preparation was done by low temperature MBE at a substrate temperature of $230^{\circ} C$. The growth rate and the Ga/Mn fraction were determined from RHEED oscillations. The (GaMn)As layers were typically 1000 \AA\  thick and the Mn content was chosen to be 5 \%. Since this was obtained from RHEED oscillations it only reflects Mn atoms incorporated in GaAs lattice sites. X-ray diffraction data for such samples has also shown that the material has high crystallographic quality \cite{sadowski}.

\begin{figure}
\resizebox{65mm}{!}{\includegraphics{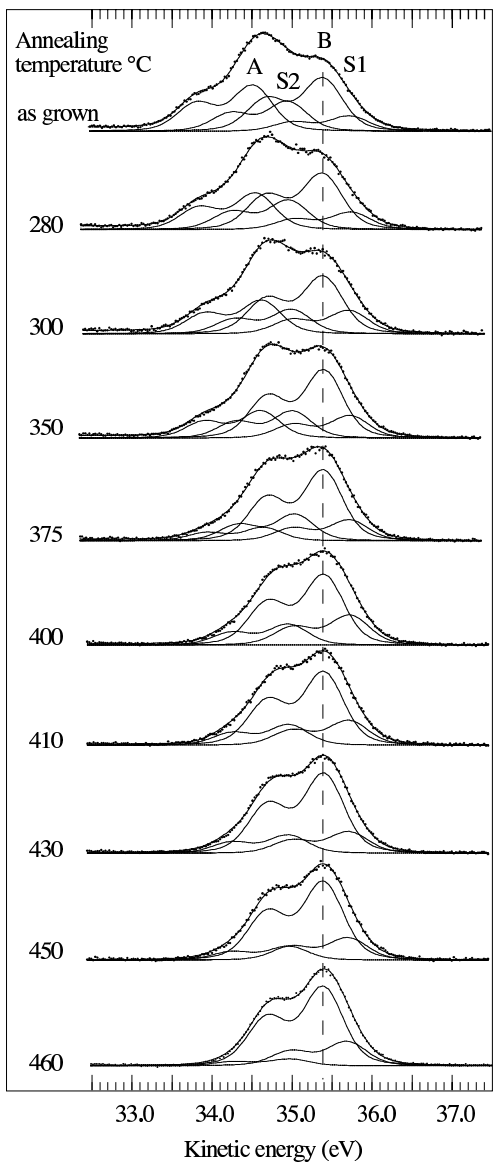}}
\caption{ Normal  emission core level spectra of As3d in
Ga$_{0.95}$Mn$_{0.05}$As for different stages in the sequential annealing process. All of the spectra were excited by 81eV photons. The spectra are aligned in energy to show the consistent behavior of the different components.}
\end{figure}

Two sets of annealing data are reported here. In the first set a single layer was annealed at sequentially increased temperatures. The time of the annealing treatments was 10 minutes at each temperature. In the second set three different samples were annealed at different temperatures for one hour each. These samples were then characterized with respect to magnetization using a superconducting quantum interference device (SQUID). For all of the samples valence band spectra and As3d spectra were recorded after each annealing. In each case the Fermi energy was determined from a Ta-foil in contact with the sample.
\\
One of the most important characteristics of semiconductor materials is the energy difference between the valence band maximum (VBM) and the Fermi level ($E_F$). In p-type materials it reflects the activation energy of charge carriers (holes), which are known to mediate the Mn spin ordering in (GaMn)As. Since it is quite difficult to extract the precise VBM  position from valence band spectra, we used the bulk component in the As3d core level spectrum as a probe of the electrostatic potential. Assuming that the energy difference between the VBM and the As3d is constant, the As3d shifts directly reflect changes in the energy difference  between $E_F$ and VBM. In this context it is important to stress that the effects of surface Fermi level pinning, normally dominant in clean semiconductors, can be safely ignored due to the very high density of bulk defects \cite{Asklund2}. To identify the bulk component in the As3d spectrum we applied a numerical spectral fitting procedure on a wide set of data. In figure 1 a set of As3d spectra is shown together with their decompositions. The decompositions are basically the same as for low temperature grown GaAs (LT-GaAs) \cite{Asklund}. Prior to any annealing the spectrum contains four components. These can be ascribed to As atoms in the bulk (B), atoms in the last layer of the As-terminated crystal (S1 and S2), and  in an As adlayer adsorbed after growth (A). The relative intensities of the different components change systematically with annealing. With an initial reduction and removal of component A, the surface reconstruction changed from (1x1) to c(4x4) typical for the As rich GaAs(100) surface. Further annealing resulted in a c(2x8) structure, also well known for As terminated GaAs(100)
\\
In Figure 2 we have plotted the binding energy of the As3d$_{3/2}$ bulk component relative to the Fermi level. We see that initially this binding energy is reduced on annealing. As argued above, the As3d shift reflects the local electronic potential, and can therefor be transferred to corresponding shifts of the VBM. Thus figure 2 shows that the Fermi level is lowered towards the VBM. This systematic lowering is observed for annealing temperatures up to around $350^{\circ}C$. Annealing at still higher temperatures results in the opposite effect - the Fermi level shifts up, away from the VBM. It is clear that the electronic properties are changed by post growth annealing even at temperatures comparable to those applied during growth.

\begin{figure}
\resizebox{75mm}{!}{\includegraphics{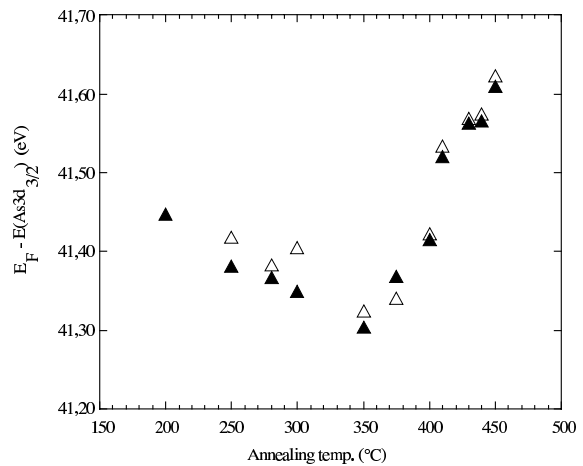}}
\caption{Fermi level relative As3d as a function of annealing 
temperature. Open symbols representing enforced fitting with rigidly fixed 
separations between different components and  filled symbols free optimum 
fittings.}
\end{figure}

Having established that the Fermi level, and thus also the hole density, is changed by annealing it is desirable to investigate the relation between the Fermi level position and magnetic properties. For this purpose three samples were prepared using the same growth parameters ($230^{\circ}C$ substrate temperature and a Mn concentration of 5$\%$) as for the sample discussed above. These samples were annealed at different temperatures and the Fermi level position was determined as previously. The Curie temperatures for the differently annealed samples were then determined  \emph{ex situ} using a SQUID. The $T_c$ values of our as-grown samples are similar to those reported in K. C. Ku et al. \cite{Ku} for layers with the same thickness, 1000\AA. The annealed samples show lower $T_c$ than what might be expect. This is due to the fact that the annealing procedure was not optimized for this purpose and also because our annealings had to be performed under UHV conditions. The highest $T_c$ values are reached for samples annealed in atmospheric conditions. The photoemission results are fully consistent with those just discussed. The Fermi level is lowered by the first annealing and is then raised after the larger "thermal dose". Although the data set is admittedly limited, the results show a clear correlation between $T_c$ and $E_F$: thus the minimum position for the Fermi energy corresponds to a maximum value for the Curie temperature. This is reasonable considering that a low Fermi energy should give a higher density of holes in the valence band \cite{Dietl}, which in turn could be expected to more effectively couple the Mn spins resulting in a higher $T_c$ value \cite{Edmonds}.

\begin{figure}
\rotatebox{270}{\resizebox{!}{75mm}{\includegraphics{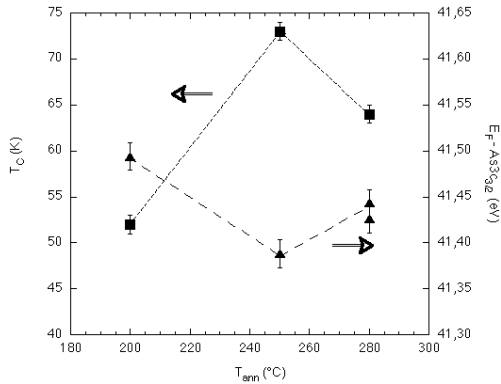}}}
\caption{Fermi energy and Curie temperature for three samples 
annealed at different temperatures.}
\end{figure}

When comparing the results of annealing (GaMn)As presented in figure 2 and in figure 3 it is apparent that the minimum for the Fermi energy occurs at different annealing temperatures. This is due to the fact that the annealing procedures were different in the two cases. We have not carried out any systematic studies of this effect, but it appears that  a similar effect can be achieved by either  high temperature \cite{Hayashi} or by long annealing time \cite{Potashnik}. It is rather the "thermal dose" than a certain temperature that should be considered within the range of conditions used here.

\begin{figure}
\resizebox{65mm}{!}{\includegraphics{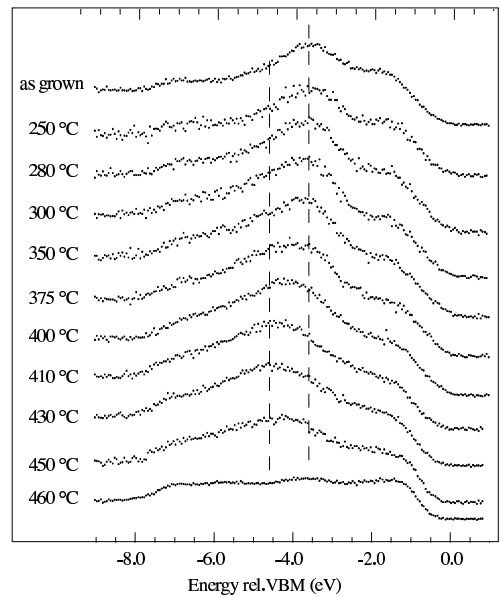}}
\caption{Normal emission valence band spectra from Ga$_{0.95}$Mn$_{0.05}$As at 
different stages of annealing. Photon energy 81 eV.}
\end{figure}

For the sequentially heated sample we also recorded valence band spectra after each annealing. These spectra were recorded in the direction of the surface normal and the photoelectrons were excited with 81 eV photons. As argued in a previous study \cite{Asklund2} the cross section of Mn3d for this photon energy is large compared to that of the GaAs valence band and thus the main peak in the VB spectra shown in figure 4 can safely be associated with Mn. After the first annealing, at $250^{\circ}C$, a slight general increase in intensity is observed. This can be ascribed to desorption of the As adlayer. Upon the following annealings the spectral shapes are rather stable, though it is clearly seen that the Mn3d induced peak around 4 eV below $E_F$ is gradually modified. More precisely, its initial position is 3.7 eV below $E_F$, but after the $430^{\circ}C$ annealing the peak occurs at 4.7 eV. Annealing at temperatures above $460^{\circ}C$ results in a dramatic reduction of this peak. Comparing figures 2 and 4 one can see that the relative reduction of the Mn3d peak at 3.7eV binding energy coincides with the upward shift of $E_F$. The minimum position of the Fermi energy correlates with the maximum value of $T_c$, which suggests that the peak at 3.7 eV is associated with Mn atoms in substitutional positions - only these atoms participate in the ferromagnetic coupling. Upon annealing the relative amplitude of this emission remains approximately constant, but its width increases such that the peak position apparently shifts towards higher binding energy. The integrated Mn3d intensity is thus increased. This increase must be ascribed to partial outdiffusion of Mn to the surface. In this kinetic energy range of 50-100 eV the spectral contribution of a surface layer is typically 30$\%$ of the total signal. We can thus estimate that the amount of segregated Mn to be around 5$\%$ of a monolayer. The conclusion of Mn segregation is consistent with recent studies of (GaMn)As capped with thin GaAs layers. It is worth mentioning that for such samples we do not see  the Mn3d peak shift toward higher binding energy after annealing \cite{adell}.

\begin{figure}
\resizebox{65mm}{!}{\includegraphics{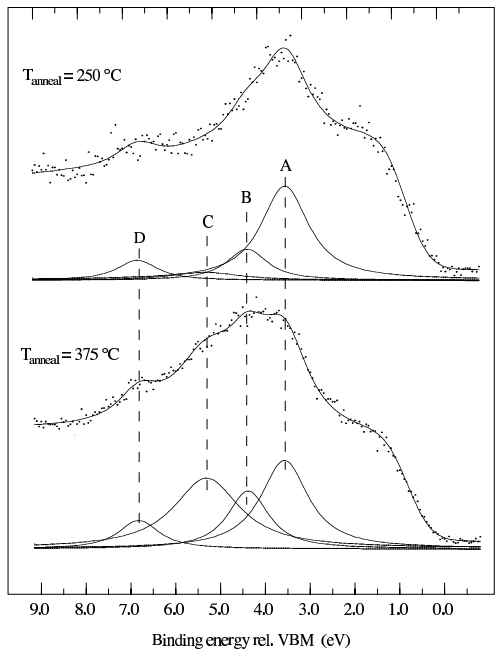}}
\caption{A simplified model has been fitted to the experimental data for the VB spectra excited by 81eV photons.}
\end{figure}

To get a better understanding of the annealing induced spectral changes we have applied the same peak fitting routine as for the As3d spectra. Since valence band spectra in general are not composed of a set of symmetrical peaks, this analysis is not expected to give a detailed physical interpretation of the observed features, but only to indicate the qualitative nature of the spectral modifications. In this analysis the valence band edge was described by a broadened Fermi function superimposed on an exponential background, and the rest of the spectrum was fitted with a set of Voigt functions. Two restrictions were applied in the fitting process: the number of peaks should be as small as possible and the peak energies should be the same in all spectra. Four peaks were needed to account for the general shape of the spectra, see figure 5. Peak A dominates the spectrum from the untreated sample and as already mentioned we associate this peak with Mn3d states in substitutional sites. We can also interpret peak D as emission from the $X_3$ critical point in the bulk Brillouin zone. Peak C is clearly enhanced with annealing and is ascribed to surface segregated Mn. The intermediate component B appears to be quite constant throughout the set of spectra. At this point we cannot suggest any explanation for this peak. There are, of course, a number of inequivalent  Mn sites which could generate a separate peak, e.g. Mn interstitials  or Mn in substitutional sites in the surface plane. In any case, the peak is certainly due to Mn, as the LT-GaAs spectrum is quite structureless in this energy range \cite{Asklund}. The shift observed in figure 4 can thus be interpreted as a result of a superposition of at least two spectral components.
\\
Returning to figure 4, we see that the overall Mn3d intensity starts to fall above  $400^{\circ} C$ annealing temperature. As mentioned above, the intensity of peak A is clearly reduced above  $350^{\circ} C$, but at annealing temperatures above  $430^{\circ} C$ the low energy component is also reduced. The most dramatic change occurs above  $450^{\circ} C$, when the low energy component is practically eliminated. Even though the Mn vapor pressure is low at these temperatures \cite{angtryck}, we believe that this reduction is due to evaporation of the surface Mn. The reduction of peak A may partly be due to outdiffusion and evaporation of substitutional Mn, but it may also be caused by phase separation and formation of MnAs clusters \cite{Boeck}. 
\\
The results presented here show that electronic properties of (GaMn)As are sensitive to thermal treatment. We find systematic changes in the energy separation between the Fermi level and the valence band maximum, and observe that the highest Curie temperature is obtained when this energy separation is small. We also find that the Mn3d induced valence band peak appears to shift almost 1 eV towards higher binding energies after the sample has been annealed. According to the presented analysis the shift should rather be described as a result of superposition of at least three peaks, whose relative intensities change with annealing. These modifications indicate that there is some structural change in the (GaMn)As samples as they are annealed. The peaks are tentatively associated to Mn atoms in substitutional bulk sites and surface segregated atoms. This sensitivity to thermal exposure underlines the importance of performing the experiment \emph{in situ}, i.e. avoiding surface preparation such as sputtering and annealing.
\\
We are pleased to acknowledge the technical support of the MAX-lab staff. This work was supported by the Swedish Research Council (VR), the Swedish Research Council for Engineering Sciences (TFR), and via co-operation with the Nanometer Structure Consortium in Lund, the Swedish Foundation for Strategic Research (SSR). One of the authors (J.S) acknowledges the financial support of the Polish State Committee of Scientific Research (KBN) through grant No: 2 PO3B05423.

\bibliography{RefAnngamnas3}
\bibliographystyle{unsrt}

\end{document}